\documentclass[aps,prb,twocolumn,showkeys]{revtex4-1}
\usepackage{graphics,graphicx,epsfig,color,hyperref,latexsym,float,amsmath}
\usepackage{times}
\input pdfcolor.tex

\begin{document}

\author{Nyayabanta Swain$^{(1,2)}$, Rajarshi Tiwari$^{(3)}$  
and Pinaki Majumdar$^{(1,2)}$}

\affiliation{
(1)~Harish-Chandra Research Institute,
  Chhatnag Road, Jhusi, Allahabad 211019, India\\
(2)~Homi Bhabha National Institute,  Training School Complex, Anushakti Nagar,
     Mumbai 400085, India\\
(3)~School of Physics and CRANN, Trinity College, Dublin 2, Ireland}

\title{Mott-Hubbard transition and spin-liquid state on the pyrochlore lattice}

\date{\today}

\begin{abstract}
The pyrochlore lattice involves corner sharing tetrahedra and the resulting 
geometric frustration is believed to suppress any antiferromagnetic order 
for Mott insulators on this structure. There are nevertheless short-range 
correlations which could be vital near the Mott-Hubbard insulator-metal 
transition. We use a static auxiliary-field-based Monte Carlo to study this 
problem in real space on reasonably large lattices. The method reduces to 
unrestricted Hartree-Fock at zero temperature but captures the key magnetic 
fluctuations at finite temperature. Our results reveal that increasing 
interaction  drives the non magnetic (semi) metal to a `spin disordered' 
metal with small local moments, at some critical coupling, and then, through 
a small pseudogap window, to a large moment, gapped, Mott insulating phase 
at a larger coupling. The spin disordered metal has a finite residual resistivity 
which grows with interaction strength, diverging at the upper coupling. We 
present the resistivity, optical conductivity, and density of states across 
the metal-insulator transition and for varying temperature. These results 
set the stage for the more complex cases of Mott transition in the pyrochlore 
iridates and molybdates.  
\end{abstract}

\maketitle

\section{Introduction}

The presence of geometric frustration disfavors 
long-range antiferromagnetic order and  promotes a complex 
magnetic state in insulating 
magnets with short-range interaction 
\cite{frust-rev1,frust-rev2}.
Two complications arise in 
magnetic insulators close to a Mott
insulator-metal transition (IMT): 
(i)~the `virtual hopping' of the 
electrons mediate long-range and multispin coupling, and
(ii)~the {\it magnitude} of the local moments weaken as the
system is pushed towards the IMT.
The first effect can lift the degeneracy of the short 
range model and promote an ordered state
while the second tends to destroy the magnetic state altogether. 
The outcome of this interplay is lattice specific
\cite{frust-Tr-lattice,frust-Kagome-lattice}, 
of relevance to several real life materials, 
and requires tools beyond those usually applied in frustrated
magnetism.

The possibilities in charge  transport are also interesting.
While the large moment, gapped, Mott phase is insulating, the
strong suppression of the moment near the IMT, and possible
orientational randomness due to frustration,  can generate
a `bad metal' state on the disordered magnetic 
background. Such a state can involve a pseudogap,
an unusually large low temperature resistivity
\cite{148-res-elm,pyr-Ir-ch-dyn,pyr-spin-ice-res}, 
and, possibly, an anomalous Hall response 
if the moments organize in a non coplanar manner
\cite{pyr-Ir-an_Hall}.  
 
The pyrochlores are a fascinating structure 
\cite{pyr-rmp} to explore these effects.
In the deep Mott state on a pyrochlore lattice, where one
expects only nearest neighbor antiferromagnetic coupling, the
effective model can be written as a sum, over the tetrahedra,
of squares of the total moment in each tetrahedron
\cite{Reimers,Moessner_Chalker}.
The  minimum of this is infinitely degenerate since the four 
spins at the vertices of each tetrahedron just need to satisfy
a zero vector sum.  The appearance of
longer range couplings as the electron-electron Hubbard
repulsion reduces (or the bandwidth increases)
can, potentially, lift the degeneracy and promote some ordered
state. 
The transport and spectral character between the spin liquid
Mott insulator and the band semimetal is also not known.

There are, remarkably, experimental realisations of 
Mott physics on the pyrochlore structure as observed 
in the rare-earth molybdates \cite{pyr-Mo-sf1,pyr-Mo-sf2,
Mo-Tc-Tf,pyr-Mo-MIT,Mo-MIT-pressure,hanasaki-andmott}
and iridates \cite{pyr-Ir-chem-pr,pyr-Ir-chem-pr2,
pyr-Ir-hydro-pr,pyr-Ir-hydro-pr2}. 
These, however, involve 4d or 5d electrons and require 
Hund's coupling or spin-orbit interaction, in 
addition to the Hubbard interaction, for a successful
description. 
To set the stage for these more realistic but complex models 
this paper focuses on the 
role of just the Hubbard interaction on the 
half filled pyrochlore lattice. 
We address two broad questions:
(i)~What is the nature of the magnetic state as one moves
towards weaker interaction in the Mott insulator, and in
particular what are the magnetic correlations near the IMT?
(ii)~What is the impact of these magnetic correlations on
electron physics (resistivity,
optics, spectral features) near the IMT?

Our main results, obtained by using a real-space static 
auxiliary-field-based method, are the following: 
(i)~Increasing interaction in the ground state
leads successively to a small-moment metal, 
then a narrow insulating pseudogap window, 
and finally to the gapped Mott insulator. 
(ii)~The resistivity and the low energy density of states have
a {\it strongly non monotonic} temperature dependence 
near the metal-insulator transition (MIT).
(iii)~The finite
moment phases, from near the MIT to the deep Mott regime,
{\it are all disordered}. Well in the Mott phase they
display what seems (within the limits of our system size) 
to be power law spatial correlations that survive to a small 
finite temperature.

The paper is organized as follows. In the next section we provide
a summary of existing results on the pyrochlore structure
- in particular the Heisenberg and Hubbard models. This
is followed by a description of our method. The next section
presents results on the phase diagram, 
density of states, transport,
and optics, across the MIT. The final section expands on a few
issues like effective models for the magnetism, computational
issues, and the magnetic structure factor.

\section{Previous work}

The pyrochlore Heisenberg antiferromagnet model, 
which describes the deep Mott limit, 
has been well studied, the Hubbard model 
much less
so. We summarize the key results from these and then
move to our method and results.

\subsection{Pyrochlore Heisenberg model}

While the Hubbard model at half-filling and 
large interaction maps on to the $S=1/2$ Heisenberg
antiferromagnet, we provide a more general discussion 
of this model below.

The nearest-neighbor {\it classical} 
Heisenberg antiferromagnet, 
$H = J \sum_{\langle ij \rangle} {\bf S}_i.{\bf S}_j,~J > 0$,
on the pyrochlore lattice 
does not show any magnetic order
down to zero temperature \cite{Reimers,Moessner_Chalker}
($T=0$).
The model can be written as $H \equiv (J/2)\sum_{\alpha} 
{\bf P}_{\alpha}^2$, upto a constant, 
where ${\bf P}_{\alpha} = \sum_{i=1}^4 
{\bf S}_i$, where the ${\bf S}_i$
are spins on the tetrahedron $\alpha$.
The ground state needs to satisfy the constraint
${\bf P}_{\alpha} = 0$ for all $\alpha$.
This results in an infinitely degenerate manifold
of possible configurations $\{ {\bf S}_i \}$.
It has been established that there are 
no internal energy barriers between these 
degenerate minima at $T=0$, 
and no free-energy barriers at finite $T$, precluding 
a freezing transition either \cite{Moessner_Chalker}.
 Since the moments are
corner shared between tetrahedra, the constraints lead
to a correlated spin structure.
The `disordered' but long-range correlated state
generates characteristic 
features in the magnetic structure factor,
appearing as ``bow ties" \cite{Zinkin} 
or ``pinch-points" \cite{Henley1},
indicative of power-law correlations \cite{Henley1,Isakov}.

In the {\it semiclassical}, $S \gg 1$, case 
the classical degeneracy is partially lifted 
by the zero-point energy of quantum fluctuations
at harmonic order, 
but there remains an infinite manifold of degenerate
collinear ground states \cite{Henley2}.
Further quantum fluctuations at anharmonic order
break the degeneracy between the various harmonic ground
states, yet they leave out a massive but nonextensive degeneracy
(smaller than the harmonic ground state) \cite{Henley3}.

The ground state for $S=1/2$ is argued to be a
quantum spin liquid \cite{Canals_Lacroix,qHAF_Balents,Nussinov}.
This spin-singlet ground state has a 
finite energy gap for triplet excitations.   
The spin-spin correlation function decays exponentially 
with distance with a correlation length shorter than 
the lattice spacing \cite{Canals_Lacroix}.

\subsection{Pyrochlore Hubbard model}

Heisenberg interactions beyond nearest neighbor, 
induce transitions to various ordered phases 
(e.g., collinear, nematic and multiple-${\bf q}$ order)
\cite{pyr_NNN_FM_Tsuneishi,pyr_NNN_FM_Chern,pyr_NNN_FM_Kuwamura}. 
Also, easy axis anisotropy, long-rage dipolar interaction,
{\it etc.}, 
lead to multiple-${\bf q}$ ordered phases
\cite{pyr_FNE_Gingras}.
It has been argued that
beyond the Heisenberg limit
the half-filled Hubbard model on the pyrochlore lattice
can be expressed as a highly frustrated 
intratetrahedral spin model with weak 
intertetrahedral
perturbations \cite{Nussinov}. 
This model has an exactly solvable Klein point, 
about which the ground state is a three-dimensional 
quantum spin liquid over an extended parameter region. 
This spin-liquid state hosts massive spinon excitations, 
which are deconfined and move in all three dimensions 
within the lattice \cite{Nussinov}.
For the Hubbard model on the pyrochlore lattice,
the only work on the Mott transition, that 
we know of, 
suggests a transition from a semimetal 
to a spin liquid Mott insulator \cite{Fujimoto}.
However, detailed properties near the IMT are not available.

\subsection{Hubbard with additional interactions}

In the pyrochlore iridates R$_2$Ir$_2$O$_7$,
both the R (rare-earth or Y) and the Ir live in two
interpenetrating pyrochlore structure. Their physics is mainly
dictated by the 5d electrons of Ir, which have strong
spin-orbit coupling and moderate Hubbard repulsion $U$
(due to large spatial extent of 5d orbitals).
This strong spin-orbit coupling lifts
the orbital degeneracy of 5d electrons,
reduces the bandwidth and leads to
an effective single band description in
terms of pseudo-spin $j_{eff} = 1/2$ states.
In the absence of $U$, the ground state is a semimetal or
topological insulator depending on the ratio of spin-orbit coupling
and hopping \cite{iridate_HF-MF} $(t)$.
This picture remains unchanged for weak $U/t$.
For strong $U/t$, Hartree-Fock calculation shows the system 
becoming an ``all-in-all-out" magnetic insulator.
Near the magnetic transition, a topological Weyl semimetal 
phase shows up\cite{iridate_HF-MF}.
A more elaborate cluster dynamical mean-field theory (C-DMFT) 
calculation confirms this scenario \cite{iridate_CDMFT}.

In the pyrochlore molybdates R$_2$Mo$_2$O$_7$,
the physics is mainly governed by the 4d electrons of Mo.
The presence of a trigonal crystal field and strong Hund's coupling
leads to band narrowing \cite{molybdate-DFT1}.
Electron correlation then acts on
a background involving 
double exchange ferromagnetism competing with
superexchange antiferromagnetism.
A Hartree-Fock calculation shows the possible
spin and orbital order in the ground state for the molybdates
\cite{molyb_HF-MF}.

\section{Model and method}

We study the single band Hubbard model, with nearest neighbor hopping,
 on the pyrochlore lattice:
\begin{equation}
H= H_0 + U\sum_{i}n_{i\uparrow}n_{i\downarrow}
\end{equation}
where $ H_0 = \sum_{ij,\sigma}
(t_{ij} - \mu \delta_{ij}) c^{\dagger}_{i\sigma}c_{j\sigma} $.
The $t_{ij}=-t$ for nearest neighbor hopping on the pyrochlore
lattice and $U >0$ is the Hubbard repulsion. 
We will set $t=1$. The chemical potential $\mu$ is varied to
maintain the density at $n=1$ as the interaction 
and temperature $T$ are varied.

We use a Hubbard-Stratonovich (HS)
transformation \cite{hubb-strat}
that introduces a vector field ${\bf m}_i(\tau)$ and a scalar
field $\phi_i(\tau)$ at each site to decouple the interaction.
This decomposition \cite{hubbard,schulz} retains the rotation
invariance of the Hubbard model, and hence the correct
low energy excitations,  and reproduces unrestricted
Hartree-Fock theory at $T=0$.

We treat the ${\bf m}_i$ and $\phi_i$
as classical fields, {\it i.e}, neglect their time dependence,
but completely retain the {\it thermal fluctuations} in
${\bf m}_i$.  $\phi_i$ is treated at the saddle point level, 
{\it i.e}, $\phi_i \rightarrow \langle \phi_i \rangle = (U/2)
\langle \langle n_i \rangle \rangle = U/2$ at half-filling, since
charge fluctuations would be penalised at temperatures $T \ll U$.
Retaining the spatial fluctuations of ${\bf m}_i$
allows us to estimate $T_c$ scales, and access the crucial thermal
effects on transport.
In the literature the overall scheme is known as 
the `static path approximation' (SPA)
to the functional integral for the partition function  
\cite{SPA_dagotto,SPA_meir}.
We have used it in the past to address the Mott transition on the 
triangular \cite{mott-Tr} lattice.  
Others have used it successfully in superconductors \cite{SPA_meir}, 
{\it etc}.
We will discuss the limitations of this scheme later in the paper.

Within this approach the half-filled Hubbard problem is 
mapped on to electrons coupled to the field ${\bf m}_i$,
which itself follows a distribution function
$P\{ {\bf m}_i \}$.  
\begin{eqnarray}
H_{eff} ~~& = &~ H_0 
- \frac{U}{2}\sum_{i}{\bf m}_{i}.\vec{\sigma}_{i}
+ \frac{U}{4}\sum_{i}{\bf m}_{i}^{2} \cr
~~\cr
P\{{\bf m}_i\} & \propto  &~
\textrm{Tr}_{cc^{\dagger}} e^{-\beta H_{eff}}
\end{eqnarray}

where the chemical potential $\mu$ in $H_0$ is
replaced by ${\tilde\mu} = \mu - \frac U2$.
$H_{eff}$ can be seen as comprising of an electronic Hamiltonian,
$H_{el}$ (the first two terms) and the classical `stiffness'
$H_{cl}= \frac U4\sum_i {\bf m}_i^2$.
In an exact calculation, where the dynamics of ${\bf m}_i$ and
$\phi_i$ are retained, $H_{eff}$ would be replaced by an
effective action while $P$ would be replaced by a fermion
determinant in the $\{{\bf m}, \phi\}$ background.

Within SPA $H_{eff}$ and $P\{{\bf m}_i\} $ 
define a coupled fermion-local moment problem. 
If the moments are large and random the electronic
problem requires numerical diagonalisation. Similarly, the
$P\{{\bf m}_i\} $ cannot be written down in closed form
since the fermion free-energy is not known for arbitrary
$\{{\bf m}_i\}$ background. The method of choice in these
situations is a combination of Monte Carlo (MC) for updating
the ${\bf m}_i$ with exact diagonalisation (ED)
of the fermion Hamiltonian
for computing the Metropolis update cost.

To access large sizes within limited time, we
use a cluster algorithm for estimating the update cost.
The energy cost of updating the variable ${\bf m}_i$ 
is computed by
diagonalizing a cluster (of size $N_c$, say) constructed 
around the site ${\bf R}_i$. 
We have extensively benchmarked this 
method \cite{tca}. 
Results in this paper are obtained for 
lattices of $6^3$ unit cells with
$4$ atoms per unit cell, using a cluster with
$3^3 \times 4$ atoms.

Electronic properties are calculated by diagonalising $H_{el}$ 
on the full lattice for equilibrium $\{{\bf m}_i\}$ configurations.
From the equilibrium configurations
we compute the single-particle density of states (DOS) as
$N(\omega) = 
\frac{1}{2N} \langle \sum_{n}\delta (\omega - \epsilon_{n}) \rangle $
where $\langle...\rangle$ indicates 
thermal average over equilibrium configurations.
For a $ L \times L \times L$ pyrochlore lattice, 
the total number of lattice sites $N = 4L^3$.
We calculate the optical conductivity
by using the Kubo formula \cite{Kubo_allen} as follows,
\begin{eqnarray}
\sigma^{xx}(\omega)&=&\frac{\sigma_{0}}{N}\sum_{n,m}
{ {f(\epsilon_{n})- f(\epsilon_{m})} 
\over {\epsilon_{m}-\epsilon_{n}} } 
\vert J^{nm}_x\vert^2 
  \delta(\omega-(\epsilon_{m}-\epsilon_{n}))
\nonumber
\end{eqnarray}
where $J^{nm}_x =  \langle n|J_x|m\rangle$ and 
$J_x$ is the current operator, given by, 
$$
J_x=-i\sum_{i,\sigma}\left
[(t c^{\dagger}_{i,\sigma}c_{i+\hat{x},\sigma}-\textrm{H.c.})\right]
$$
$f(\epsilon_{n})$ is the Fermi function,
$\epsilon_{n}$ and $|n\rangle$ are 
the single-particle eigenvalues and eigenstates of
$H_{el}\{{\bf m}_i\}$ respectively.
The conductivity is in units of
$\sigma_{0} = e^2/(\hbar a_0)$, 
where $a_0$ is the lattice constant.
The d.c. conductivity is obtained as a low frequency average of the
optical conductivity over a window $\sim 0.05t$.
We calculate average local magnetic moment
$m_{avg} = \frac{1}{N} \sum_{i}^{N} \langle |{\bf m}_i| \rangle$
and their distribution 
$P(m) = \frac{1}{N} \langle \sum_{i} \delta (m - |{\bf m_{i}}|)\rangle $
from the equilibrium configurations.
The magnetic structure factor is computed as
$S({\bf q}) = \frac{1}{N^2}\sum_{ij}\langle{\bf m}_i.
{\bf m}_j\rangle e^{i{\bf q}.({\bf r}_i-{\bf r}_j)}$
at each temperature.
The rapid growth of a mode at some ${\bf q} = {\bf Q}$ (say) 
indicates the onset of magnetic order.
Our results are averaged over 100 equilibrium Monte Carlo configurations. 

\begin{figure}[t]
\centerline{
\includegraphics[width=8.0cm,height=7.0cm]{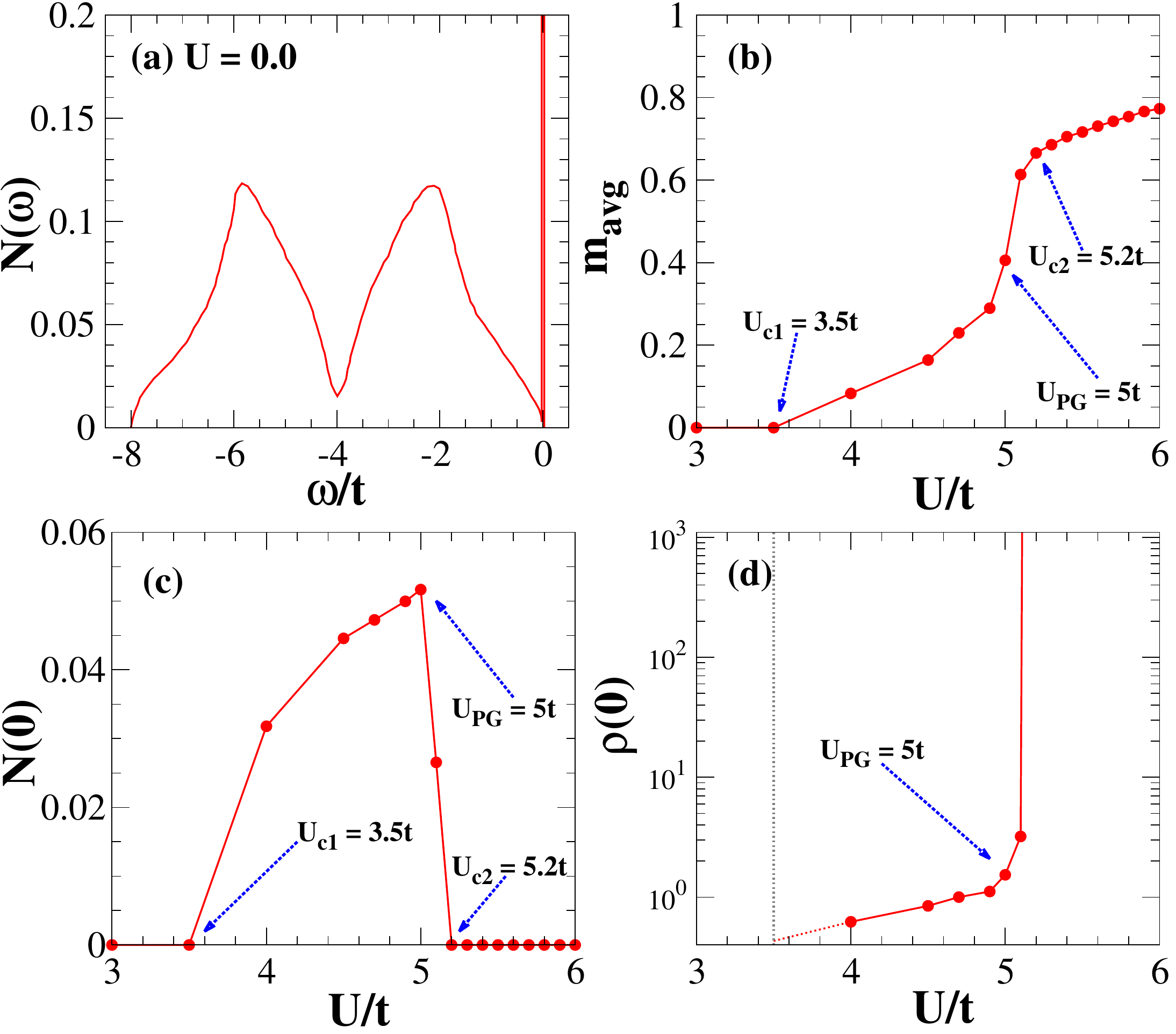}
}
\caption{\label{zt}Colour online:
(a)~Tight binding density of states for the pyrochlore lattice.
(b)-(d) Results at $T=0$ for: 
(b)~variation of the average local moment $m_{avg}$,
(c)~the density of states, $N(0)$, at the Fermi level,
and (d)~the resistivity $\rho(T=0)$.
} 
\end{figure}

\section{Results}

\subsection{The ground state}

As $T \rightarrow 0$ our MC mainly samples configurations that
maximise $P\{{\bf m}_i\}$, or, alternately, minimize the energy,
{\it i.e.}, 
$\frac{\delta}{\delta{\bf m}_{i}} \langle H_{eff} \rangle = 0$,
This is the same
as unrestricted Hartree-Fock in the magnetic channel.

Upto a critical coupling $U_{c1} \sim 3.5t$ the minimisation 
yields $m_{i} = \vert {\bf m}_i \vert = 0$ at all sites. 
As a result upto $U_{c1}$ the
electronic ground state is essentially tight binding, the
density of states for which is shown in Fig.\ref{zt}.(a).  
There is a sharply suppressed DOS at Fermi level 
characteristic of the pyrochlore band structure, 
and a flat band right above the Fermi level. 
For $U < U_{c1}$ the system is a semimetal.

For $U_{c1} < U < U_{PG}$ where $U_{PG} \approx 5t$ 
we observe a small moment, orientationally disordered, 
magnetic state. The average moment size is shown in
Fig.\ref{zt}.(b), and the full distribution later in the paper.
With the disorder caused by these moments
breaking the translation invariance of the pyrochlore lattice,
the DOS at the Fermi level,
$N(0) = \int_{- \Omega}^{\Omega} N(\omega)d \omega / \int_{- \Omega}^{\Omega} d \omega $,
where $\Omega=0.05t$,
gains weight (Fig.\ref{zt}.(c)). 
In a narrow region around $U_{PG}$ 
there is rapid increase in the mean magnitude of the moments and, 
as a result, the DOS at Fermi level gets depressed again. 
The detailed behavior of the low energy DOS is shown later.
At $U_{c2} \sim 5.2t$ the DOS at the Fermi level vanishes as
a Mott gap opens. For $U > U_{c2}$ the moments are large,
 and saturate to their atomic value, $\vert {\bf m}_i \vert=1$, 
as $U/t \rightarrow \infty$. 
The coupling of the electrons to the local moments leads to
weak scattering and a small resistivity for $U_{c1} < U < U_{PG}$,
a rapid growth in resistivity for $U_{PG} < U < U_{c2}$, and zero
d.c conductivity for $U > U_{c2}$. 
This is shown in Fig.\ref{zt}.(d). 
We will discuss the resistivity in much greater detail later, 
and just wanted to highlight 
the effect within the $T=0$ mean-field state here.

A comment about the magnetic state.
Since there was no reason to expect that the moments
would have any obvious periodic pattern the
only way to do the `minimisation' was via simulated 
annealing employing Monte Carlo. 
In the $U/t \rightarrow \infty$ limit the half filled 
Hubbard model leads to a Heisenberg model for the
${\bf m}_i$ (classical, in our
SPA approximation) and for the pyrochlore Heisenberg 
antiferromagnet the moments are
known to be disordered \cite{Reimers,Moessner_Chalker}, 
albeit power law correlated \cite{Henley1,Isakov}. 
Our MC minimization reproduces this state. At lower $U/t$,
both in the Mott phase and the `metal', there are no
simple results known - but our results suggest that a
`disordered' state persists and correlations reminiscent of
a classical spin liquid phase \cite{Henley1} 
survive down to $U/t \sim 8$.

\begin{figure}[t]
\centerline{
\includegraphics[width=4.5cm,height=4.30cm]{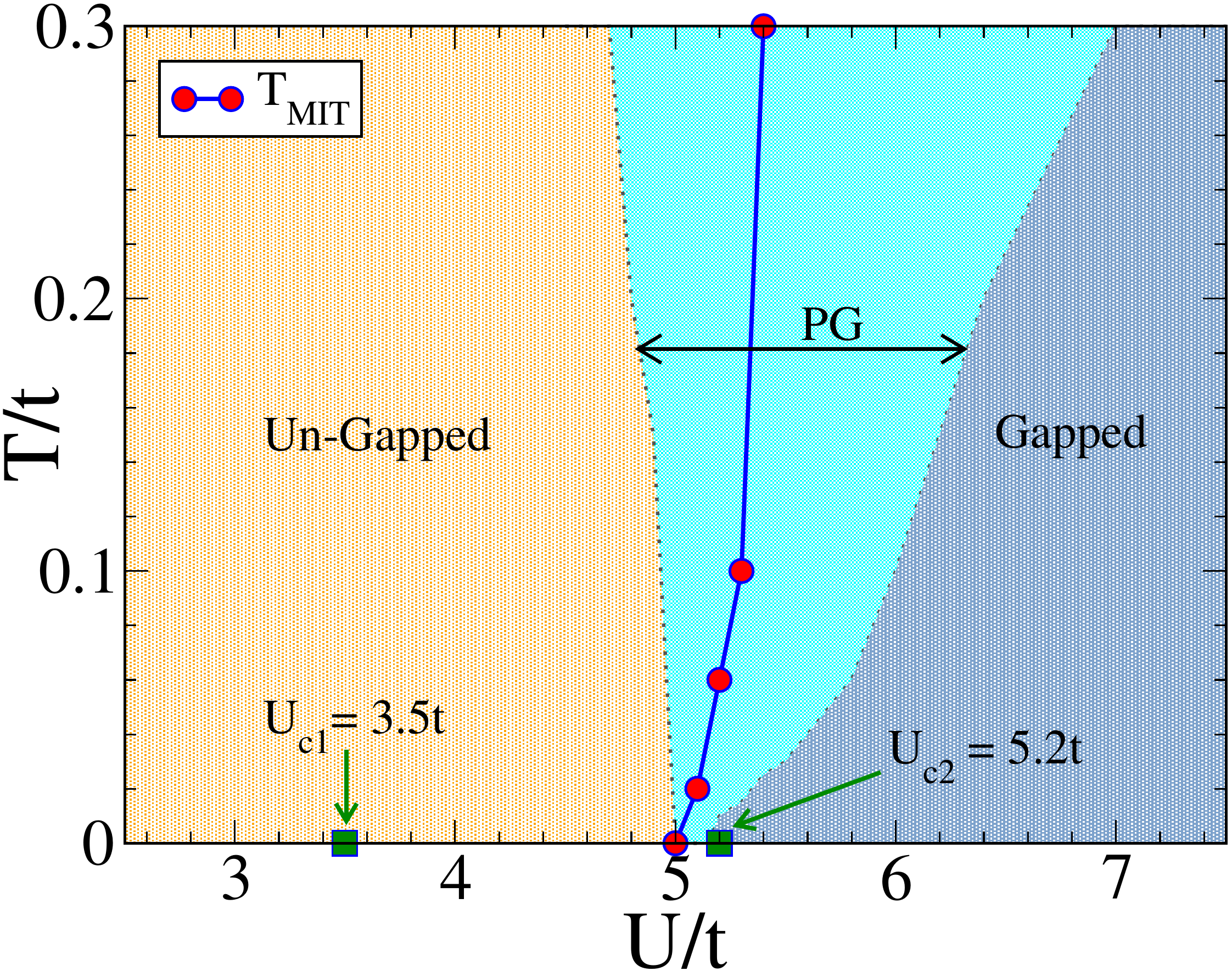}
\includegraphics[width=4.0cm,height=4.30cm]{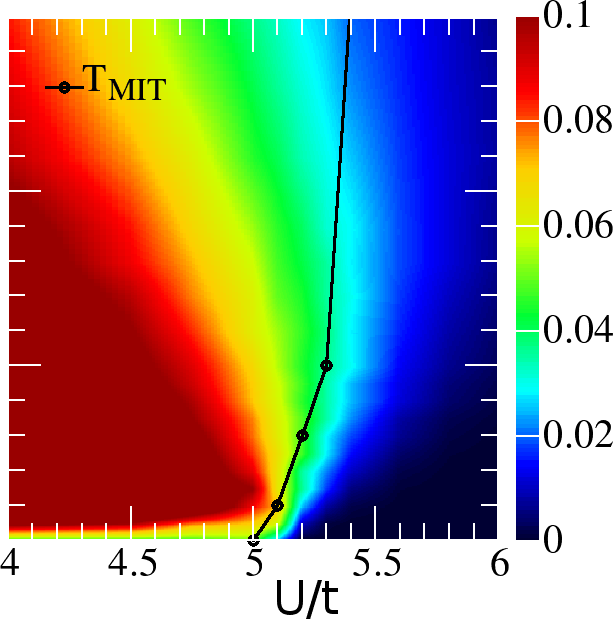}
}
\caption{\label{pd} Colour online: Phase diagram (left) and 
the density of states at the Fermi level (right) 
for varying $U$ and $T$.
Local moments appear at $U_{c1}$ but the state remains metallic,
turning insulating at $U_{c2}$.
PG refers to a pseudogap state
and the metal-insulator transition line separates regions with
opposite signs of $d\rho/dT$. The right panel highlights the
`re-entrant' feature in the low energy DOS with increasing
temperature.
To avoid clutter we 
have not marked the $U_{PG}$ scale in the ground state in
the left figure.
}
\end{figure}

\begin{figure*}[t]
\centerline{
\includegraphics[width=17.5cm,height=5.8cm]{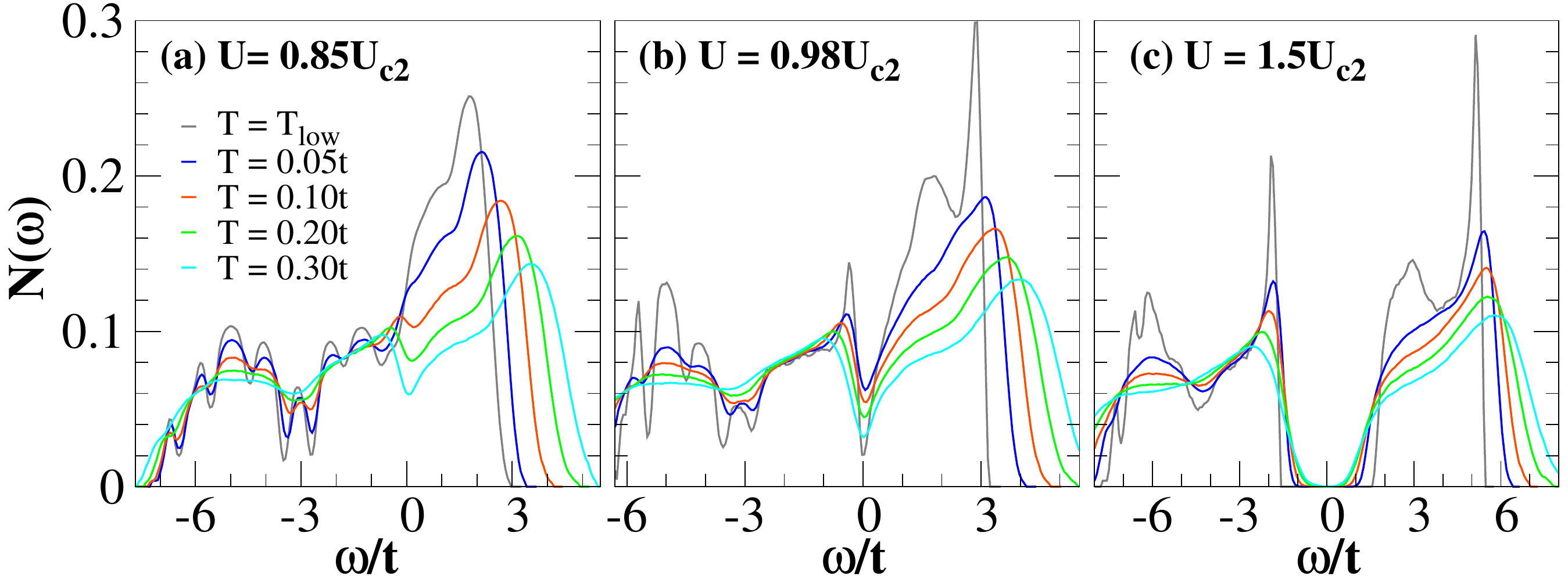}
}
\caption{\label{dos}Colour online: Density of states
varying with temperature
at three representative regimes of our calculation.
(a)~$U = 0.85U_{c2}$ lies in the gapless metallic side,
(b)~$U = 0.98U_{c2}$ corresponds to the pseudo-gap (PG) regime
with a pronounced dip in the DOS at the Fermi-level.
and (c)~$U = 1.5U_{c2}$ lies in the gapped Mott-insulating side.
$T_{low} = 0.01t$ in panel (a), where as $T_{low} = 0.0$ for panels (b) and (c).
In panel (a) the low energy DOS reduces with increasing $T$, in panel (b)
it increases and then reduces with $T$, and in panel (c) it
monotonically increases with $T$.
}
\end{figure*}

\subsection{Thermal phase diagram}

Figure \ref{pd} (left panel) shows the $U-T$ phase diagram in terms of
the magnetic, transport, and spectral properties that we observe.
The following features emerge:

At finite $T$ thermal fluctuations of local moments 
on the weak-coupling side ($U < U_{PG}$) lead to a quick
low $T$ increase 
in the low energy DOS, and then a gradual decrease with further
increase in $T$. 
On the strong-coupling side ($U > U_{c2}$) the angular fluctuations 
of the local moments result in a slight smearing of the Mott-gap with 
temperature and an increase in the low energy DOS. 
However in the Mott-transition neighborhood,
the Mott gap quickly converts to a PG with increasing
$T$, leading to the widening of the PG region shown
in Fig. \ref{pd} (left panel).

We demarcate the finite $T$ metal-insulator boundary in terms of
the temperature derivative $d\rho/dT$. A state is metallic if
$d\rho/dT > 0$ and insulating if $d\rho/dT < 0$. 
The spectral features and resistivity are discussed in detail
further on.
  
Figure \ref{pd} (right panel) shows the DOS at the Fermi level 
varying with $U$ and $T$. 
On the Mott insulating side ($U \geq U_{c2}$) we observe
the DOS slowly increasing with temperature, 
which can be understood as the filling of the Mott gap.
On the metallic side ($U_{c1} < U < U_{c2}$) 
we see a non monotonic behavior.
The DOS quickly grows with temperature in the low temperature regime.
It then reduces with further increase in temperature,
the weight getting transferred to high energy.

\subsection{Density of states}

Figure \ref{dos} shows the thermal evolution of the DOS 
in three of the four broad interaction regimes of our phase-diagram.

$(i)$ For $U<U_{c1}$ the ground state is characterized by
$m_{i} = |{\bf m}_i| = 0$. The electron
model reduces to the usual tight-binding pyrochlore lattice.
This is characterised by two flat-bands 
at the upper band edge and vanishing DOS at the Fermi energy. 
At finite $T$,  small `randomly' oriented 
local moments appear in the system broadening the flat bands
and leading to a small DOS at the Fermi level.
$(ii)$ For $U_{c1} < U < U_{PG}$, the ground state has small 
disordered local moments.
The DOS is gapless and the weight at the Fermi level is
nonmonotonic with $T$, increasing initially and then decreasing
as a weak PG forms (see Fig. \ref{dos}.(a)).
$(iii)$ For $U_{PG} < U < U_{c2}$ the DOS has a PG at $T=0$. This
fills up initially with increasing $T$, 
Fig. \ref{dos}.(b), but deepens again above
a temperature scale that is visible in Fig. \ref{pd} right panel.
$(iv)$ For $U \gg U_{c2}$ the ground state has a hard gap.
With increase in temperature, the angular fluctuations 
of the local moments result in a slight smearing 
of the Mott-gap,
and an increase in the low energy DOS
Fig. \ref{dos}.(c). However, there exists a clear 
Mott-gap until very high temperature, $T \sim m_{avg} U$.

\begin{figure}[b]
\centerline{
\includegraphics[width=6.0cm,height=7.2cm]{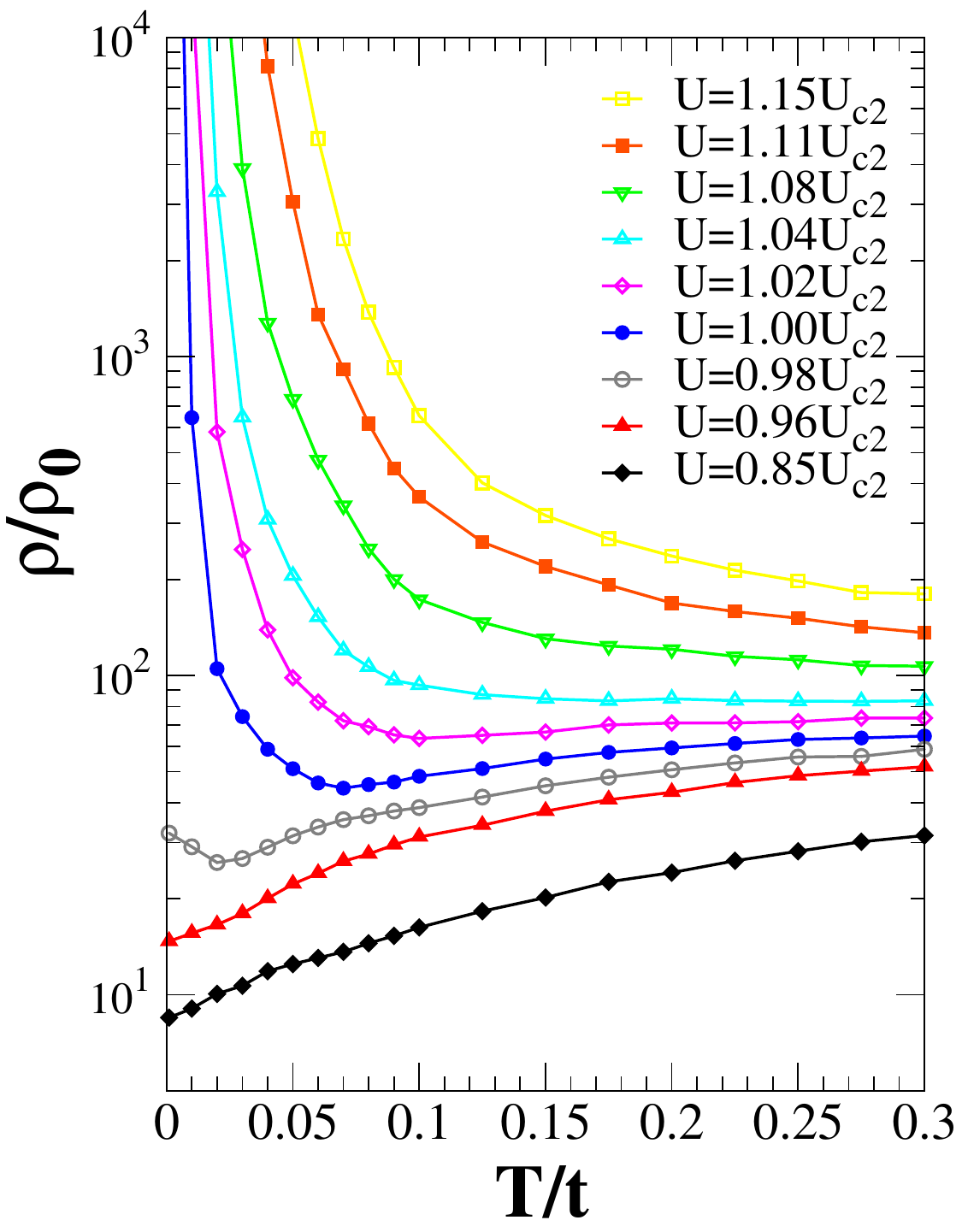}
}
\caption{\label{res}Colour online:
Temperature dependence of the resistivity for different $U/t$
near the Mott insulator-metal transition.
The normalizing scale is $\rho_{0} = \hbar/e^2$.
}
\end{figure}
\begin{figure}[t]
\centerline{
\includegraphics[width=8.7cm,height=4.4cm]{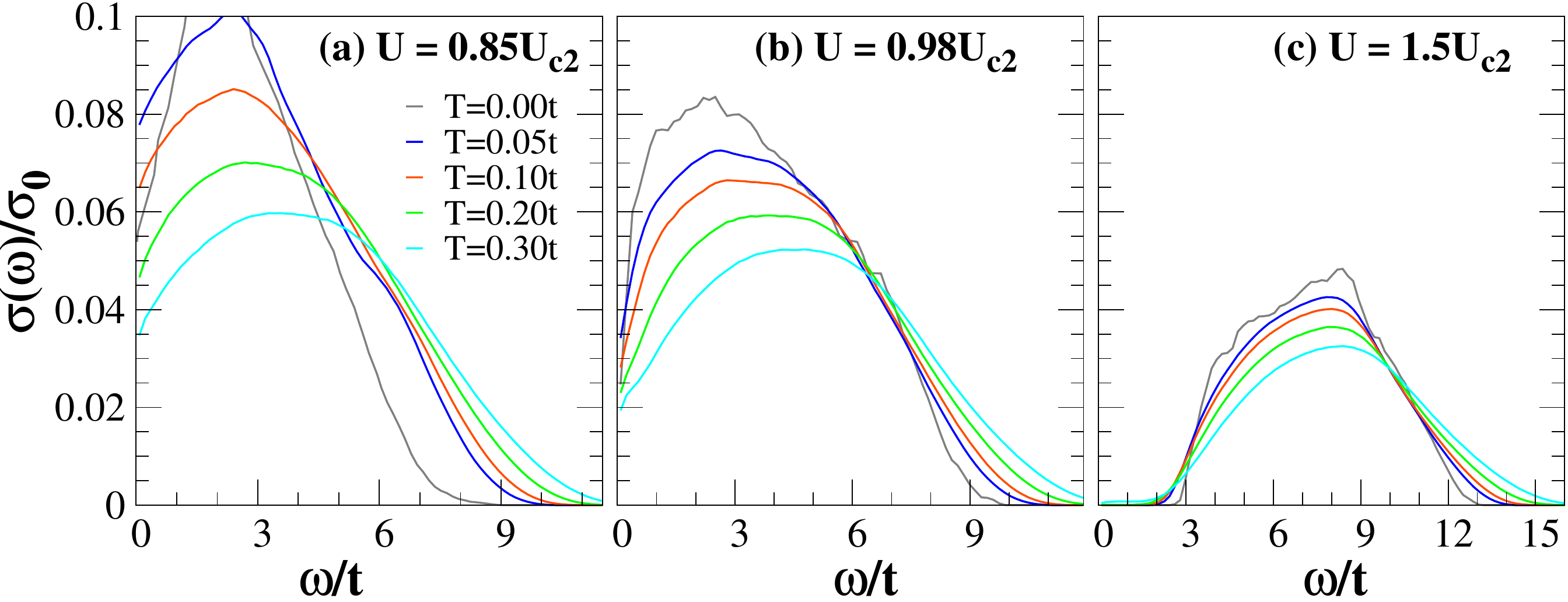}
}
\caption{\label{opt}Colour online: 
Optical conductivity
at $U/U_{c2} = $0.85, 0.98 and 1.5 with varying temperatures.
$U = 0.85U_{c2}$ shows a non-Drude like behavior with peak
at small and finite frequency. With increasing $U$
the peak moves to higher frequency and
the zero frequency weight decreases continuously and
eventually there appears a gap for $U \geq U_{c2}$.
}
\end{figure}

\subsection{Transport and optics}

Figure \ref{res} shows the d.c. resistivity $\rho(T)$
for different $U/t$. We consider the four regimes.

$(i)$~For $U < U_{c1}$ 
the $T=0$ phase is a semimetal. Since this lies
well below the Mott transition we do not show the
$T$ dependence here.
$(ii)$~For $U_{c1} < U < U_{PG}$ 
the residual resistivity $\rho(0)$ is finite
with $d\rho/dT >0$ over the entire $T$ range.
The resistivity can be understood in terms of 
a disorder induced density of states and 
the scattering of electrons from the small 
disordered moments.
This is the metallic regime.
$(iii)$~For $U \gg U_{c2}$ the system has a 
clear Mott gap at $T=0$ with $\rho(0) \rightarrow \infty$.
In this regime $d\rho/dT <0$ over the entire temperature window. 
This is the Mott-insulating regime.
$(iv)$~In the neighborhood of $U_{c2}$,
{\it i.e}, $\vert U - U_{c2} \vert \ll U_{c2}$,
~$\rho(T)$ shows a non-monotonic behavior.
We observe $d\rho/dT <0$ in the low temperature limit,
crossing over to  $d\rho/dT >0$ with increasing $T$.
The temperature at which $d\rho/dT$ changes its sign 
is indicated as the $T_{MIT}$.

We observe $T_{MIT}$ increasing with $U$ as seen in Fig.\ref{pd}.
This behavior can be understood as the scattering of electrons 
from the background fluctuating local moments. 
As $U$ increases, the average local moment magnitude 
$m_{avg}(U)$ also increases, resulting in the increased 
scattering of the electrons and a depleting DOS at the
Fermi level.

Figure \ref{opt} shows the optical conductivity from
our calculation as we cross the Mott 
transition. The important points are as follows:
(i)~$\sigma(\omega)$ for $U < U_{c1}$ is a semimetal at
$T=0$ and does not have a Drude peak.
(ii)~For $ U_{c1} <U < U_{PG}$, ~$\sigma(\omega)$  shows 
a response with the peak at a small finite frequency
that slowly shifts to higher values with increasing $T$.
(iii)~For $U > U_{c2}$ the system has a clear gap $\Delta(T)$ in the DOS. 
Thus $\sigma(\omega) = 0$ for $\omega < \omega_{c} \sim \Delta(T)$.
With increasing temperature the gap $\Delta(T)$ reduces, resulting in small, 
but increasing low frequency weight of $\sigma(\omega)$ 
and the peak position shifts to higher frequency. 
This Mott-insulating regime of the pyrochlore lattice 
may have finite spectral weight at $\omega =0$ in the
optical conductivity $\sigma(\omega)$ at high temperatures.
(iv)~For $U_{PG} < U < U_{c2}$ we have a pseudogap in the DOS.
$\sigma(\omega = 0) \rightarrow 0$ in this regime.
However with increasing temperature 
the zero frequency weight increases initially 
and then decreases in accord with the behavior of the DOS 
in this regime. 

\section{Discussion}

\subsection{Overall scenario}

Within our picture, the interaction effects are encoded in the 
presence of the `local moments' ${\bf m}_i$. The {\it size}
$m_i$ of this moment dictates the on site splitting at the
site ${\bf R}_i$, and leads to a Mott gap in the overall
DOS when 
$U m_{avg} \gg t$. The spatial correlations of ${\bf m}_i$ 
decide whether
the electron spin ${\vec \sigma}_i$ will display any long
range order.

Let us correlate the electron physics across the
Mott transition to the behavior of the ${\bf m}_i$,
we will then take up the effective models for the
${\bf m}_i$ themselves.
(i)~In the metallic regime,  $U \gtrsim U_{c1}$,
the $m_i$'s are small and
the orientations are random. The scattering from
these moments leads to a finite broadening, ${\tau}_{\bf k}^{-1}$,
of the momentum eigenstates. This generates a
finite DOS at $\omega=0$, a non-Drude
optical response, and resistivity increasing 
with temperature
(unlike in a semimetal).
(ii)~In the Mott-insulating regime, $U > U_{c2}$,
the $m_i$'s are large and show
short-range magnetic correlation (discussed next).
The large $m_i$'s lead to a Mott gap in the DOS
and optical conductivity,
and a diverging resistivity as $T \rightarrow 0$.
(iii)~In the pseudogap regime, the $m_i$'s are moderately
large and orientationally disordered.
This results in a strong suppression in the
DOS at Fermi energy, but no gap, a
large finite residual resistivity
and a non-Drude optical response.

We have shown $m_{avg}$ for varying $U$ at $T=0$
(Fig. \ref{zt}(b)).  
In the next section we discuss the limiting models
that dictate the behavior of ${\bf m}_i$, and 
in the section after, we show detailed results on 
the size distribution of ${\bf m}_i$, and its spatial
correlations.

\begin{figure*}[t]
\centerline{
\vspace{.3cm}
\includegraphics[width=13.4cm,height=4.0cm]{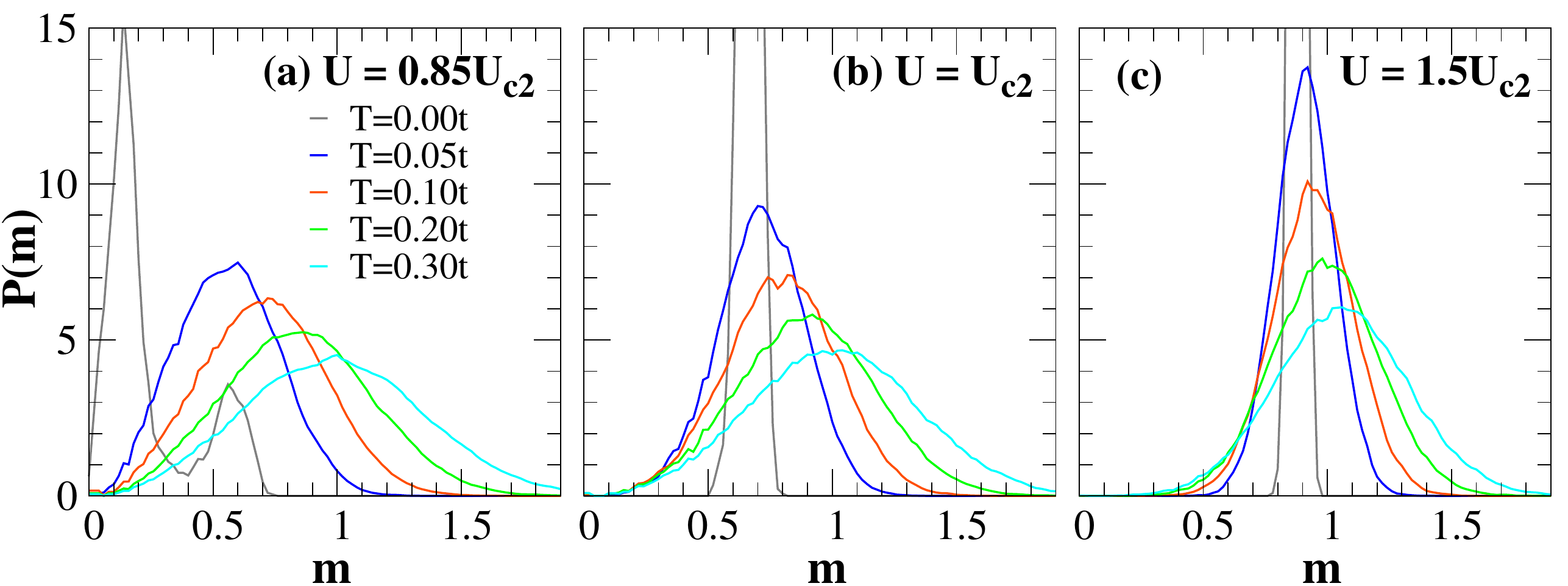}
}
\caption{\label{pm}Colour online:
Temperature dependence of $P(m)$
for $U = 0.85U_{c2}$, $U_{c2}$ and $1.5U_{c2}$
for indicated temperatures.
}
\end{figure*}

\subsection{Effective models in limiting cases}

We have shown only a formal expression for the effective
magnetic Hamiltonian. To get a feel for the magnetic states
that arise it is useful
to provide the approximate analytic structure in limiting
cases. These are (a)~weak coupling, when $U \lesssim t$,
and (b)~strong coupling, when $U \gg t$.

\subsubsection{Weak coupling}

Our effective interaction looks like a `Hund's coupling',
with the electron spin coupled to a background moment through
the coupling $U$. Given the formal similarity with the
Hund's problem we can borrow the form of the weak coupling 
result \cite{akagi_udagawa} from the literature:
\begin{eqnarray}
H_{eff}\{{\bf m}_i\}  \sim &&   
-{U^2 \over 4} \sum_{ij} 
(\chi_{ij} - {1 \over U} \delta_{ij}) {\bf m}_i.{\bf m}_j \cr
&& ~~~+ {U^4 \over 16} \sum_{ij}^{kl}
f({\bf m}_i, {\bf m}_j, {\bf m}_k, {\bf m}_l) + .. \nonumber
\end{eqnarray}
The structure of $f({\bf m}_i,...{\bf m}_l)$ is
complicated, involving two-spin terms 
such as ${\bf m}_i.{\bf m}_j$, $({\bf m}_i.{\bf m}_j)^2$, 
three-spin terms such as $({\bf m}_i.{\bf m}_j)({\bf m}_i.{\bf m}_k)$, 
and four-spin terms such as $({\bf m}_i.{\bf m}_j)({\bf m}_k. {\bf m}_l)$.
The $i,j,k,l$ can be separated by long distance in the weak-coupling limit.

In contrast to the Hund's problem, where there were predefined local
moments, in our case the moments have to arise from an instability
in the electron system. The leading instability involves the 
vanishing of coefficient of the quadratic term, {\it i.e.}, 
$ 1 - U\chi_0({\bf q}) = 0$. This generates the moment and, 
if there is a prominent peak at some ${\bf q} = {\bf Q}$ 
in $\chi_0$ the moments order with that wavevector.
In such a situation the fourth order term can be expanded
about ${\bf Q}$. The quartic term decides the
magnitude of the order at ${\bf Q}$.

We have computed the $\chi_0({\bf q})$ for the half-filled
pyrochlore band, having tested the scheme for the quarter-filled
band for which results are available \cite{pyr_Chern}.
The $\chi_0({\bf q})$ is featureless,
suggesting that there is no particular wavevector that
would be picked out.
In that case the quartic term, whose detailed structure
we do not know at half filling, 
decides not only the magnitude but also the spatial
character of the order parameter field.
It seems that 
the nonlinearity creates a bimodal distribution 
for the $m_i$, discussed further on, but without
any significant spatial correlation.

\subsubsection{Strong coupling}

For $U \gg t$, one can write an effective magnetic Hamiltonian 
on the pyrochlore lattice by tracing out fermions order 
by order in $t/U$. 
This gives us
\begin{eqnarray}
H_{eff}\{{\bf m}\} & = & H_{tetr}\{{\bf m}\} + H_{coup}\{{\bf m}\} \cr
~~ \cr
H_{tetr}\{{\bf m}\} & \sim & 
\sum_{\alpha} ( J_0 + J_2 {\bf P}_{\alpha}^2 + J_4 {\bf P}_{\alpha}^4 + ... )
~~\cr
H_{coup}\{{\bf m}\} & \sim & 
J'_4 \sum_{i\in \alpha ,j\in \beta} {\bf S}_i . {\bf S}_j 
+J''_4 \sum_{i\in \alpha ,j\in \beta}^{k\in \alpha \cap \beta} 
({\bf S}_i.{\bf S}_k)({\bf S}_j.{\bf S}_k)... \nonumber
\end{eqnarray}
where  ${\bf P}_{\alpha} = \sum_{i=1}^{4} {\bf m}^{\alpha}_{i}$
is the total spin on the tetrahedron $\alpha$, 
$H_{tetr}$ describes interactions between spins in a tetrahedron
while $H_{coup}$ includes the intertetrahedron terms
with a common corner shared site.
$J_0 = -\frac{8t^2}{U}(1 - \frac{4t^2}{U^2})$,
~$J_2 = \frac{t^2}{2U}(1 - \frac{24t^2}{U^2})$,
~$J_4 = \frac{5t^4}{8U^3}$
and $J'_4, J''_4 \sim  O(\frac{t^4}{U^3})$. 

Deep in the Mott phase, 
one would drop the $J_4$, ~$J'_4$ 
and $J''_4$ terms 
and obtain a classical Heisenberg model, which
does not show any long-range order or freezing,
but power-law correlations at low temperature
\cite{Henley1,Isakov}.
The correlations of the electron spins, ${\vec \sigma}_i$,
can be computed in response to this.
As $U \rightarrow U_{c2}$ the $J_4$, $J'_4$ 
and $J''_4$ terms become important. 
These multi-spin exchange interactions
modify the magnetic ground state, but some of the Heisenberg
limit features \cite{Henley1} are observable (see next)
in the structure factor down to $U \sim 8t$.
The expansion in $t/U$ 
ceases to be useful once the gap closes.

\begin{figure*}[ht!]
\centerline{
~~~~\includegraphics[width=13.4cm,height=14.2cm]{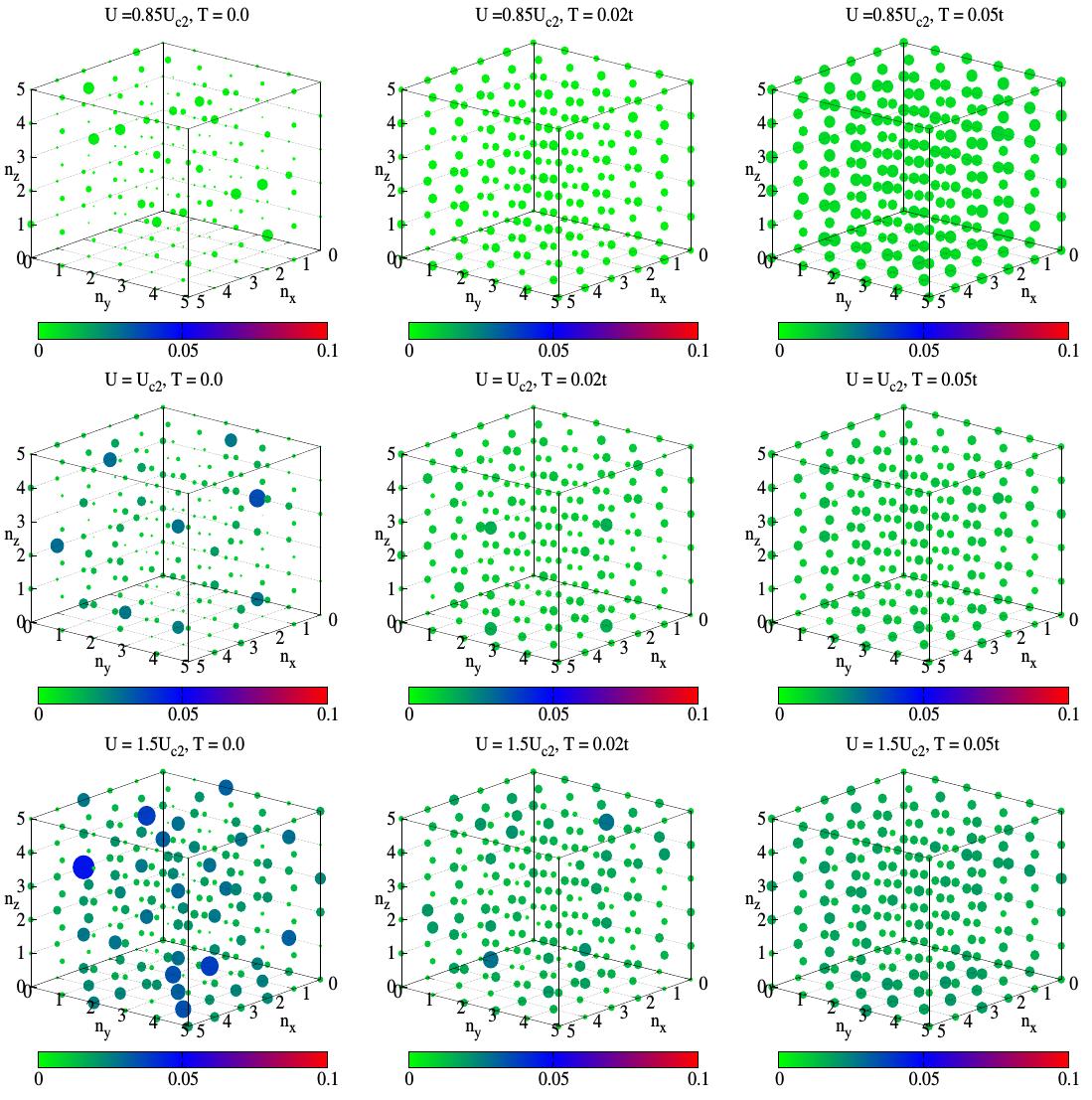}
}
\caption{\label{sf}Colour online:
The full magnetic structure factor $S({\bf q})$
for $U = 0.85U_{c2}$, $U_{c2}$, $1.5U_{c2}$ (along column)
and $T = 0.0$, $0.02t$ and $0.05t$ (along row).
We use the notation ${\bf q} = \frac{2\pi}{L}(n_x,n_y,n_z)$,
n's are integers.
The  size of a dot signifies relative weight at a given
${\bf q}$ while its color represents the actual magnitude
of $S({\bf q})$.
}
\end{figure*}

\subsection{Detailed magnetic structure}

Figure \ref{pm} shows the the amplitude distribution
$P(m)$ of the magnetic moment for different $T$ and 
interaction regimes.

Figure \ref{pm}(a) shows $P(m)$ for $U = 0.85U_{c2}$.
In the ground state there is a two peak structure, 
highlighting the presence of amplitude inhomogeneity.
We have checked that there is no significant density
inhomogeneity, or charge ordering, in the system.
With rise in temperature, $P(m)$ shows 
a broad single peak behavior with the peak shifting
towards large $m$. This behavior is seen in the 
$U_{c1} < U < U_{c2}$ regime.
Figure \ref{pm}(b) shows $P(m)$ for $U \sim U_{c2}$,
just in the Mott insulating side.
The ground state has a narrow single peak feature 
indicating an amplitude homogeneous Mott state.  
With rise in temperature this narrow peak broadens, 
the peak position moves towards higher $m $.  
Figure \ref{pm}(c) shows $P(m)$ for $U = 1.5U_{c2}$,
well in the Mott regime. $P(m)$ has a single peak feature 
which broadens with temperature and shifts to 
higher $m$. The fluctuations about the mean are 
weaker in the insulator than in  the metallic state.

Figure \ref{sf} shows the ${\bf q}$ dependence of the 
magnetic structure factor 
for varying $T$ and $U$. 
We observe that even at $T = 0$, ~$S({\bf q})$ has no ordering
peak at any ${\bf q}$'s. The magnetic ground state is disordered.
Nevertheless for both $U \sim U_{c2}$ and $U = 1.5 U_{c2}$
the weight distribution is not completely homogeneous in ${\bf q}$
and have some prominent features. 
This signature survives to $T \sim 0.03t$.
Examination of the classical Heisenberg model has
revealed that there are `pinch point' features in
scattering, arising from the constrained ground state,
that survive \cite{Conlon_Chalker} to $T \sim 0.1J$, 
where $J$ is the exchange scale (in our case $J = t^2/U$). 
Obtaining such a result,
indicative of power law correlations, requires larger
system size $(\gtrsim 10^3)$, much longer annealing
($\gtrsim 10^8)$ MC weeps), and a more sophisticated 
algorithm instead of single spin update. 
Due to our computational cost and size 
limitations we only get a hint of this spin liquid state. 
By the time $T \sim 0.05t$, which is $\sim J/2$ at $U = 1.5 U_{c2}$,
the ${\bf q}$ dependence is completely featureless.

\subsection{Methodological issues}

There are several approximations, analytic and numerical, we had
to make to achieve some headway. Let us comment on these
one by one, and their effect on the reliability of our results.

\subsubsection{Static approximation in the Monte Carlo} 

In principle both spatial and temporal fluctuations could be important
near the Mott transition on the pyrochlore lattice.
However, fully handling the temporal fluctuations of the
${\bf m}_i$ and $\phi_i$ fields requires a quantum
Monte Carlo scheme. The absence of such results is probably
due to the sign problem for fermions in the frustrated geometry.
We have retained only the zero Matsubara frequency,
$\Omega_n =0$, mode of the auxiliary fields, exactly.  

It is possible, but non trivial, to set up a 
Gaussian expansion for the finite $\Omega_n$ 
modes of ${\bf m}_i$ and $\phi_i$, while retaining and 
treating the $\Omega_n=0$ mode exactly as we have.
This is a project for the future. In the absence of
such a calculation we can only make the following 
conjectures.

(i)~Local moments would remain well defined in the insulating
phase, due to the gap. Since the present
theory itself predicts no order at any $U$ or temperature, we 
do not expect additional quantum fluctuations to
qualitatively modify the   magnetic state. 
(ii)~A possible {\it qualitative consequence} of quantum fluctuations
could be the restoration of translation invariance in
the `magnetic metal' phase. 
We cannot make a definite comment on this but
even if this were to happen, making the metallic state
perfectly conducting at $T=0$, we believe that above a low
coherence temperature one would see the signature of a highly
resistive metal.

\subsubsection{Finite size effects}

We have done our calculations on lattices that involve 
$\sim 800$ atoms.
This is much larger than what is typically accessible in 
full-fledged fermion quantum Monte Carlo (QMC) 
(since temporal fluctuations also have to be included), 
allowing us to access resistivity and spectral features
which do not involve artificial broadening.
It is possible to try the calculations on $8^3 \times 4$ lattices
(involving about $2000$ atoms) and checks at individual parameters
do not reveal a significant difference.

\begin{figure}[t]
\centerline{
\includegraphics[width=6.0cm,height=4.6cm]{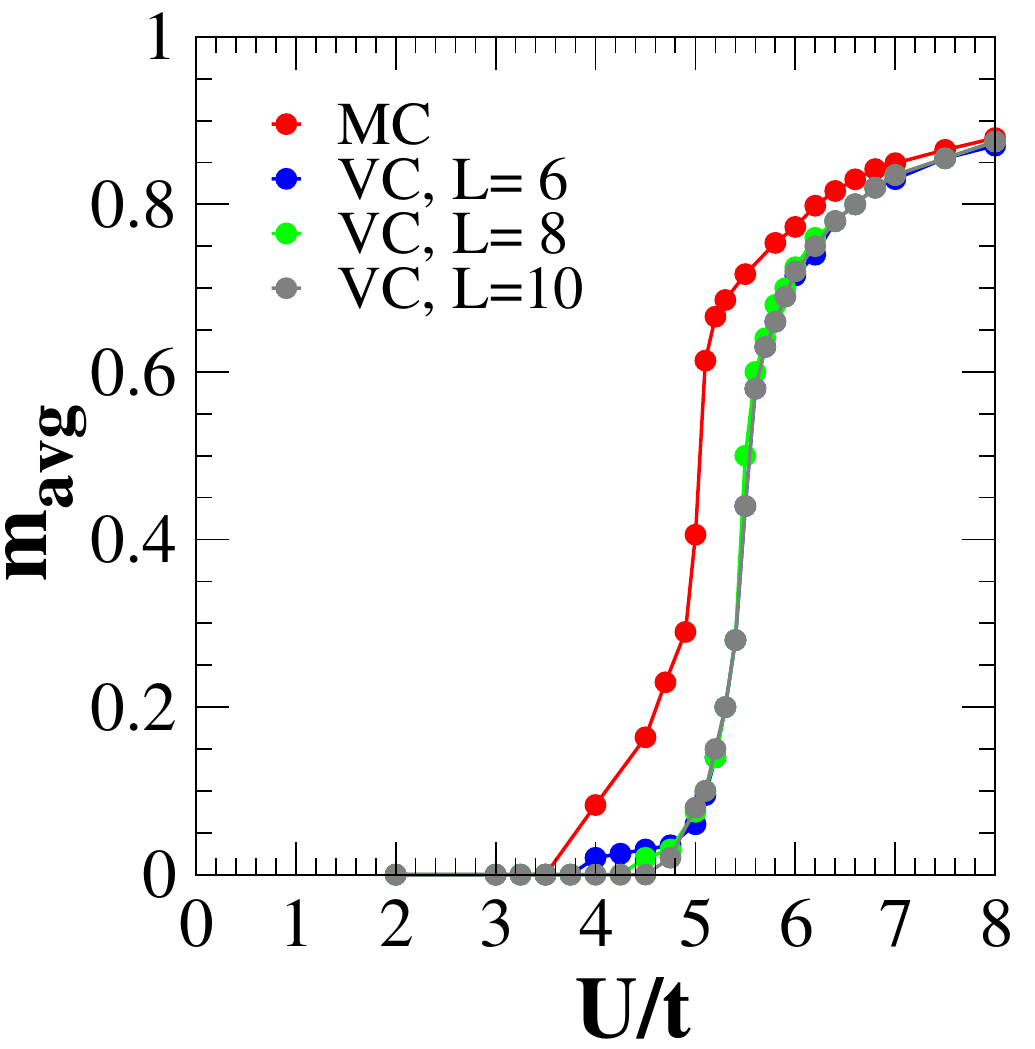}
}
\caption{\label{vc}Colour online:
Comparison of the variationally obtained average moment value
with result from the Monte Carlo as $T \rightarrow 0$.
}
\end{figure}

\subsubsection{Cluster based update}

Our Metropolis update involves a small cluster rather than 
diagonalisation of the full Hamiltonian. This is well controlled
in the large $U$ limit when the `range' of electron excursion is
limited but less reliable near the Mott transition. For that
purpose we have used a variational calculation, described below,
that uses the MC result as an ansatz and checks the stability of
such a state on large (up to $10^3 \times 4$ lattices) for the
ground state. 
The results are discussed below - and are qualitatively
consistent with the cluster based MC.

\subsubsection{Variational check}

Since MC hints that the magnetic ground state involves disordered
local moments for $U > U_{c1}$, we tried a simple variational check.
We set up trial configurations $\{ {\bf m}_i \}$ with random orientation, 
but uniform magnitude $m_0$, with $m_0$ as a variational parameter. 
This differs from the real situation 
where the ${\bf m}_i$'s have some amplitude inhomogeneity and
also orientational correlation. 
Energy minimisation confirms the presence
of a small moment phase, beyond some $U_{low}$, with an 
initial slow growth of $m_0(U)$ and then a rapid crossover to large
values at some $U_{high}$ (see Figure \ref{vc}). 
The $U_{low}$ and $U_{high}$ are about $10\%$ higher than MC estimates 
and may get reduced if spatial correlations are included.


\onecolumngrid
{\begin{center}
\begin{table}
\begin{tabular}{ | p{3.5cm} | p{4.5cm} | p{4.9cm} | p{4.4cm} | }
\hline
~~~Indicator              & ~~~Present theory      & ~~~Molybdate        & ~~~Iridate       \cr
\hline
Degrees of freedom     & One orbital per site    
                       & Two orbitals per site and a $S=1/2$ moment 
                       & Effective one orbital per site \cr
\hline
Interactions           & Repulsive Hubbard   
                       & Repulsive Hubbard, Hund's coupling, antiferro superexchange        
                       & Repulsive Hubbard, spin-orbit coupling \cr
\hline
Weak-coupling phase    & Band semimetal, spin disordered metal 
                       & Ferromagnetic metal, spin-glass metal 
                       & Correlated metal \cr
\hline
Strong-coupling phase  & Spin-liquid Mott insulator
                       & Spin-glass Mott insulator
		               & All-in-all-out Mott insulator \cr
\hline
Features in resistivity &  Continuous divergence of $\rho(0)$ with $U/t$ 
                        &  Seemingly continuous divergence of $\rho(T=0)$ with $r_R$
                        &  Seemingly continuous divergence of $\rho(T=0)$ with $r_R$  \cr
\hline
Magnetoresistance in moderate field  & Weak 
                                     & Huge: drives insulator-metal transition (Gd$_2$Mo$_2$O$_7$)
                                     & Weak \cr
\hline
Anomalous Hall response   & Not explored 
                          & Observed in Nd$_2$Mo$_2$O$_7$
                          & Observed in Nd$_2$Ir$_2$O$_7$ \cr
\hline
\end{tabular}
\caption{\label{tab} Comparison of features of the pyrochlore Hubbard model
with the models relevant for the molybdates and iridates.}
\end{table}
\end{center}
}

\twocolumngrid

\subsection{Relation to experiments}

Pyrochlore Mott materials include the
rare-earth based
molybdates R$_2$Mo$_2$O$_7$ and
iridates R$_2$Ir$_2$O$_7$ 
(R is a rare-earth ion or Y).
The Mo and Ir ions live on a pyrochlore lattice while the R
inhabit an interpenetrating pyrochlore structure.

In contrast to 3d electron
based systems which are dominated by
the Hubbard $U$, the molybdates involve 4d electrons - where
the Hund's coupling $J_H$ is also important, while the
iridates involve 5d electrons - with spin-orbit coupling
playing a vital role. In this paper we focused
on the simple Hubbard model as the starting problem. In
what follows we quickly list out the key molybdate and iridate
Mott features (see Table \ref{tab}) 
and argue how some of these are already visible
in our present results.

The molybdates exhibit a MIT with decreasing rare-earth 
ionic radius $r_A$ \cite{}. 
Materials with larger $r_A$ are ferromagnetic (FM) metals, 
those with small $r_A$ are spin glass (SG) insulators 
\cite{pyr-Mo-sf1,pyr-Mo-sf2},
and there is a SG metal phase \cite{Mo-MIT-pressure} near the MIT. 
Pressure and magnetic field can drive an IMT in 
materials like Gd$_2$Mo$_2$O$_7$
which are weakly insulating \cite{hanasaki-andmott}. 
An anomalous Hall effect (AHE) has been observed 
\cite{Mo-anm-hall1,Mo-anm-hall2,Mo-anm-hall3}
in Nd$_2$Mo$_2$O$_7$ and is ascribed to 
non vanishing spin chirality.

The iridates also show a MIT 
with decreasing $r_A$, but in this case the MI transition is
accompanied by a magnetic transition from a paramagnetic to
an antiferromagnetic `all-in-all-out' (AIAO) ordering 
\cite{pyr-Ir-chem-pr,pyr-Ir-chem-pr2,iridate_AIAO}.
While the magnetic character differs distinctly
from molybdates, iridates also show a pressure driven
insulator-metal transitions, via unusually resistive
ground states \cite{pyr-Ir-hydro-pr,pyr-Ir-hydro-pr2,pyr-Ir-ch-dyn},
and spin chirality driven AHE \cite{pyr-Ir-an_Hall}
in materials like Pr$_2$Ir$_2$O$_7$ \cite{pyr-Ir-msl}.

Reducing $r_A$ reduces the hopping in these materials and
is akin to increasing the $U/t$ ratio in our model.
The general observation
\cite{pyr-Ir-chem-pr,pyr-Ir-chem-pr2,Mo-Tc-Tf,pyr-Mo-MIT}
of a metal-insulator transition
with reducing $r_A$ is consistent with our phase
diagram. In the table above, however we
compare a set of  indicators
between molybdate or iridate 
experiments and our present theory.
While the presence of electron correlation on a pyrochlore
structure is common to all three, there are 
additional interactions present in the real materials.
The one common feature that emerges is the occurence of
a high resistivity ground state close to the MIT which,
we believe, is due to the coupling of electrons to
disordered spin and orbital moments. This is a general
consequence of the interplay of correlation effects and
strong geometric frustration. We will discuss the
detailed iridate and molybdate theories
elsewhere.

\section{Conclusion}

We have studied the Mott transition in half-filled Hubbard model 
on a pyrochlore lattice. The geometric frustration and 
the corresponding large magnetic degeneracy prevents 
the occurence of any magnetic order in the deep Mott state. 
This continues all the way to the insulator-metal
transition. Beyond the insulator-metal transition 
there is a window with a pseudogap in the density of states, 
disordered local moments, and a large residual resistivity. 
At even weaker interaction one recovers the non magnetic 
band semimetal. Thermal fluctuations destroy 
the `spin-liquid' correlations in the insulating state, 
converting the system to an uncorrelated paramagnet. The low energy 
electronic density of states and the resistivity show a monotonic
temperature dependence deep in the metallic and insulating phases,
but a non monotonic character near the insulator-metal transition.
A comparison with the rare-earth molybdates and iridates reveal
a clear similarity in the resistivity, 
while detailed aspects of the magnetic state
and magnetotransport are missed out. 
We are exploring these problems
separately.

\begin{acknowledgments}

We acknowledge use of the HPC clusters at HRI and a 
discussion with Shubhro Bhattacharjee. 
We thank Sauri Bhattacharyya for a reading of the manuscript. 
PM acknowledges support from an Outstanding Research Investigator 
grant of the Department of Atomic Energy-Science Research Council 
(DAE-SRC) of India.

\end{acknowledgments}

\bibliographystyle{unsrt}

\end{document}